\documentclass{elsart}

\usepackage{natbib}
\usepackage{graphicx}% Include figure files
\usepackage{dcolumn}% Align table columns on decimal point

\usepackage{amssymb}
\newcommand{\su}{\uparrow}
\newcommand{\gu}{\downarrow}

\begin{document}

\begin{frontmatter}


 \title{Effective electron-electron and electron-phonon interactions in the Hubbard-Holstein model}
%\thanksref{label1}}
 \author{G. Aprea,}
 \author{C. Di Castro,}
 \author{M. Grilli\corauthref{cor1},}
 \ead{marco.grilli@roma1.infn.it}
 \corauth[cor1]{M. Grilli}
\author{and J. Lorenzana}
  \address{INFM-CNR
 SMC Center, and Dipartimento di Fisica\\
Universit\`a di Roma "La Sapienza", piazzale Aldo Moro 5, I-00185 Roma, Italy}   %\thanksref{label3}}

%%\title{}


%%\author{}

%%\address{}

\begin{abstract}
% Text of abstract
We investigate the interplay between the electron-electron and the electron-phonon 
interaction in the Hubbard-Holstein model. We implement the flow-equation
method to  investigate within this model the effect of correlation on 
the electron-phonon effective coupling and, conversely, the effect of phonons in the 
effective electron-electron interaction. Using this technique we obtain analytical 
momentum-dependent expressions for the effective couplings and we study their behavior
for different physical regimes. In agreement 
with other works on this subject, we find that the electron-electron attraction
mediated by phonons in the presence of Hubbard repulsion is peaked at low transferred momenta.
The role of the characteristic energies involved is also analyzed.
\end{abstract}


\end{frontmatter}

% main text
The electron-phonon ({\it e-ph}) interaction can play an important role in
many physical systems, where it can be responsible for instabilities of
the metallic state leading, e.g., to charge density waves and superconductivity.
When the {\it e-ph} is particularly strong  polaronic effects can also
be responsible for a huge increase in the effective electronic masses.
On the other hand the issue of strong electron-electron ({\it e-e}) correlations
is of relevance for many solid state systems like, e.g., (doped) Mott insulators, 
heavy fermions, and high-temperature superconductors. Therefore the 
interplay between the {\it e-e} correlation and the {\it e-ph} coupling
is far from being academic and its understanding is of obvious pertinence 
in several physically interesting cases. In particular, in the superconducting cuprates the
{\it e-e} correlations are strong and are customarily related to  the marked
anomalies of the metallic state, including the linear temperature dependence 
of the resistivity\cite{fry}, \cite{gunnarson}, where phonons are not apparent.
This led to believe that the {\it e-ph} interaction plays a secondary role
in these materials, and that the main physics is ruled by electron correlation. 
However, the presence of polaronic spectroscopic features  \cite{polaronsincuprates}
as well as various  isotopic effects \cite{isotope} in underdoped superconducting cuprates
shows that the {\it e-ph } coupling is sizable in these systems. 
Recent results obtained with Angle Resolved Photo-Emission Spectroscopy 
(ARPES) \cite{lnzr} have also been interpreted in terms of electrons
substantially coupled to phononic modes. These contradictory 
evidences naturally raise the issue of why correlations can reduce the
{\it e-ph} coupling in some cases and not in other. In this regard it has been
often advocated \cite{kimtesanovic,grilli,zeyher,scalapino} that 
strong electron correlation suppresses the {\it e-ph} large angles scattering,
which mainly enters in transport properties, much more than the forward scattering,
which is instead relevant in determining charge instabilities \cite{grilli}.
Therefore the momentum dependence of the {\it e-ph} interaction can give rise to
important effects and select relevant screening processes due to correlations.

In this paper we study the effect of (a weak) electronic correlation on a two 
dimensional Holstein {\it e-ph} system using the flow-equation technique developed
by F. Wegner \cite{wgnr1,wegner2}. The flow-equations describe the
evolution of the parameters of the Hamiltonian under an infinite
series of infinitesimal canonical transformations labeled by a
parameter $l$ that rules the flow. The generator of
the canonical transformation is chosen in such a way to eliminate some
interactions in favor of other interactions. Shall one solve the 
flow equations exactly the initial hamiltonian at $l=0$ and the final
hamiltonian  at $l=\infty$ are guaranteed to have the same eigenvalues.
 Of course in general the flow-equations are not solvable, but suitable
approximations can be introduced to obtain qualitative trends. 
Under certain approximations one can map the interacting problem 
into a noninteracting one.  In this case the method can be 
roughly seen as a way to derive a Fermi liquid theory where bare
fermions evolve into quasiparticles under the flow.
Since one can choose what parts of the hamiltonian remain, the method 
can also be exploited to get effective interactions.
Of course, in getting a fully diagonal Hamiltonian one is naturally led to 
discover the ``low-energy'' phsysics of the problem. 
This is the approach (and the spirit of it) previously adopted 
by Wegner and coworkers to analyze the instabilities of various forms of the
Hubbard model \cite{grt,hkvych,hankevych}.  Another application of the
method is the derivation of effective interactions. For example  
this scheme was used to obtain
an effective {\it e-e} interaction by eliminating phononic degrees of freedom in
the absence of the initial {\it e-e} interaction \cite{lnz}.
Rather than obtaining the low-energy physics one
obtains an interacting Hamiltonian which still has to be
solved. Analyzing the  structure of the resulting effective
interaction we will discuss the physics of the {\it  e-ph}
interaction in the presence of the  {\it e-e} interaction and
vice versa. We will see that
this procedure provides the correct energy scales for the phonons dressing
the effective {\it e-e} interaction and, conversely, of the {\it e-e} interaction
dressing the {\it e-ph} interaction.

The paper is organized as follow: first, in Sec. \ref{s_draft_2}, 
we introduce the Hubbard-Holstein
Hamiltonian and, in Sec. \ref{s_draft_3}, we present the formalism of the 
flow equations technique. Then in Sec.
\ref{ss_draft_4_1} we validate the method by applying it to a two atom system.
In Sec. \ref{ss_draft_4_2} we
compute the effective {\it e-ph} interaction in a two-dimensional system and in Sec. \ref{s_draft_5}
we eliminate the {\it e-ph} interaction to get an effective {\it e-e} coupling, thus
providing a model Hamiltonian suitable to discuss different physical regimes.
Finally the summary and the conclusions are provided in Sec.(\ref{s_draft_6}).

%%%%%%%%%%%%%%%%%%%%%%%%%%%%%%%%%%%%%%%%%%%%%%%%%%%%%%%%%%%%%%%%%%%%%%%%%%%%%%%%%%%%%%%%%%

\section{Model}           \label{s_draft_2}
\begin{figure}[!ht]
\begin{center}
\includegraphics[scale=.50,angle=90]{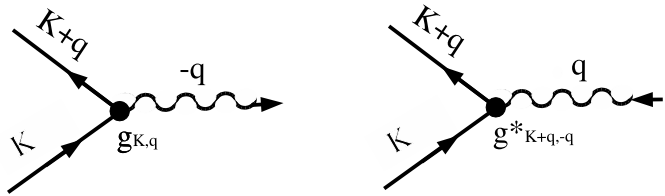}
\caption{Bare {\it e-ph} vertex. Streight lines represent electron propagators while wavy
lines are for phonons.  With our notation, in the vertex with an outgoing phonon the first index
of $g$ represents the incoming electron momentum, while in the vertex with incoming phonon
it represents the outgoing electron momentum; in both cases the second index of $g$
represents the opposite of the phonon momentum.}
\label{e-ph_vertex}
\end{center}
\end{figure}

We consider a system of interacting electrons and phonons described by the 
Hubbard-Holstein model, where the  electron density is locally coupled with
a dispersionless phonon. For the sake of simplicity and since this case
is relevant in physical systems of broad interest like the cuprates and the
manganites, we consider the model in a two-dimensional lattice
\begin{equation}
H =  \sum_{ i,j,\sigma} t_{i,j} c^{\dagger}_{i,\sigma} c_{j,\sigma}
-\mu\sum_{i\sigma} n_{i\sigma}
+ U \sum_i n_{i\uparrow} n_{i\downarrow} 
 + g_0\sum_i n_i(a_i +a^{\dagger}_i) + \omega_0 \sum_i a^{\dagger}_i a_i,
\label{hamiltonian}
\end{equation}
where $n_i\equiv n_{i\uparrow}+n_{i\downarrow}$, with $n_{i\sigma}\equiv 
 c^{\dagger}_{i,\sigma} c_{i,\sigma}$. 
$a$($c$) and $a^{\dag}$($c^{\dag}$) are the phonon (electron) 
annihilation and creation operators. For simplicity we consider only the
nearest-neighbor hopping $t_{i,j}=-t$ for $i$ and $j$
 nearest-neighbor lattice sites. The above Hamiltonian defines the
 starting point of our flow.  Since under the flow new interactions will be
 generated, in order to describe the evolution of the Hubbard-Holstein parameters, 
we have to write the  Hamiltonian in a more general form
\begin{eqnarray}                                     \label{e_draft_0.5}
H &=&  T + V + G \\
T &=& \sum_{q} \omega_q :a^{\dag}_{q} a_{q}: + 
\sum_{k,\tau} (\epsilon_k-\mu) :c^{\dag \tau}_{k} c^{\tau}_{k}:  \\
V &=& \frac{1}{2N}\sum_{k,k',q,\tau} V_{k,k',q}
:c^{\dag \tau}_{k+q} c^{\dag -\tau}_{k'-q} c^{-\tau}_{k'} c^{\tau}_{k}:  \\
G &=& \frac{1}{\sqrt{N}}\sum_{k,q,\tau}(g_{k,q} \ a^{\dag}_{-q} + g_{k+q,-q}^* \ a_{q}) 
:c^{\dag \tau}_{k+q} c^{\tau}_{k}:
\end{eqnarray}
Figure \ref{e-ph_vertex} shows the diagrammatic representation of the
{\it el-ph} interaction. In the following we assume the coupling $g_{k,q}$ to be real, while
$\omega_q$ and $\epsilon_k$ represent the phonon and electron dispersions  respectively. 
 $V_{k,k',q}$ is the {\it e-e} interaction. $N$ is the number of lattice 
sites. Colons indicate normal ordering of the operators with respect to the Fermi sea.
The expression (\ref{e_draft_0.5})represents the Hamiltonian at a generic value of the flow 
parameter $l$ whereas Eq.~\ref{hamiltonian} provides the initial
values of the parameters at $l=0$, i.e.  $g_{k,q}(l=0)=g_0$, $V_{k,k',q}(l=0)=U$,
$\omega_q(l=0)=\omega_0$, and $\epsilon_k=-2t(\cos(k_x)+\cos(k_y))$ (here we use a
unit lattice spacing).

%%%%%%%%%%%%%%%%%%%%%%%%%%%%%%%%%%%%%%%%%%%%%%%%%%%%%%%%%%%%%%%%%%%%%%%%%%%%%%%%%%%

\section{Method}                     \label{s_draft_3}
The flow-equation method \cite{wgnr1,wegner2} is based on an algebraic technique, 
the Jacobi method, to (block)
diagonalize Hamiltonians. A series of infinitesimal rotations is performed
such that at each step the non-diagonal elements get smaller while the others 
asymptotically approach a finite
limit. If $l$ is the parameter ruling the flow, this technique performs a unitary
transformation $U(l)$ such 
that $H(l)=U(l)H(l=0)U^{\dag}(l)$. One can absorb all the evolution in
the parameters of the Hamiltonian generating a new $H(l=\infty)$. This transformed
Hamiltonian has the property that, when expressed in terms of the original operators,
it has the desired diagonal (or block-diagonal) form with
renormalized couplings. The spectrum naturally stays unchanged. 
The evolution of the Hamiltonian is controlled by 
\begin{equation}               \label{e_draft_0.55}
\frac{dH(l)}{dl}=[\eta(l),H(l)],
\end{equation}
where the anti-hermitian generator $\eta(l)=\frac{dU(l)}{dl}U^{\dag}(l)=-\eta^{\dag}(l)$.
Proper identification of the numerical coefficients in this operator equation yields the 
flow equations. In a finite Hilbert space, one can use a matrix representation of the
$H$ and $\eta$ operators bringing Eq. \ref{e_draft_0.55} to a closed system of equations.
On the other hand, working in second quantization, in general the commutator in Eq.
\ref{e_draft_0.55} will generate $N$-body operators, which are not necessarily
included in the starting Hamiltonian. In this case a truncation is needed at some
$N$ in order to obtain a closed form. For operators of conserved
particles, this truncation is naturally justified in the low-density limit.
On the other hand, for phonons, this truncation is only allowed
within a weak-coupling approximation.

The first step is to find a generator that produces the convergence of
the flow to the desired form of the Hamiltonian. We split the Hamiltonian 
in $H_0$ plus a part that has to be eliminated $V$.  
Wegner has shown that the following  choice produces the desired flow \cite{wgnr1,wegner2}
\begin{equation}                                                                     
\label{e_draft_0.75}
\eta (l)=[H_0(l),V(l)]=[H_0(l),H(l)]=[H(l),V(l)].
\end{equation}
This method has been used to eliminate one kind of interaction
(electron-phonon) in favor of 
an other effective (electron-electron) interaction \cite{lnz}. The method was also
adopted to select a specific channel in the electron-electron
Hubbard interaction to see how this  is dressed by all other channels in 
order to study the different possible instabilities of the model\cite{grt,hkvych,hankevych}. 
Here, as a first step starting from the model Hamiltonian
(\ref{e_draft_0.5}), in the weak correlation limit, we use this technique 
to eliminate the {\it e-e}
interaction and study how the resulting effective {\it e-ph} interaction 
behaves. Since this use is somehow unconventional we will first test 
if the method is suitable to
this purpose in a toy model with two sites and a truncated phonon Fock space and 
then proceed to the general case. Subsequently we will consider the 
(more standard \cite{lnz}) opposite procedure to eliminate the {\it e-ph}
interaction in favor of an effective {\it e-e} interaction.

%%%%%%%%%%%%%%%%%%%%%%%%%%%%%%%%%%%%%%%%%%%%%%%%%%%%%%%%%%%%%%%%%%%%%%%%%%%%%%%%%%%%%%%%%
\section{Effective Electron-Phonon interaction}                     \label{s_draft_4}

\subsection{Two-site lattice}                           \label{ss_draft_4_1}                           
\label{ss_draft_1}

As mentioned above, our first aim is to eliminate the {\it e-e} 
interaction term dressing the {\it e-ph} one. In order to achieve this result we use the generator
\begin{equation}                \label{e_draft_0.8}
\eta(l)=[T+G,V].
\end{equation}
In this case we are not properly trying to diagonalize the Hamiltonian but we want to 
eliminate some off-diagonal elements corresponding to the {\it e-e} interaction to see how 
other off-diagonal elements, the ones corresponding to the {\it e-ph} interaction, 
renormalize. 
Thus it is necessary to
check if the generator given above yields a solution with vanishing $V_{k,k',q}$ and finite
effective coupling $g_{k,q}$.
An analytical proof is beyond the purposes of this paper and therefore we will proceed 
constructively by analyzing a two-site Hubbard-Holstein model 
as a first non-trivial playground to apply this technique. 

One first remark is in order: The in-phase ionic
displacement corresponds to a rigid shift of the two-site ``lattice'' and
decouples from the electronic degrees of freedom. Therefore in momentum space the
phonons at zero momentum can be disregarded and we only will consider phonons
at momentum $\pi$. Accordingly only the {\it e-ph} couplings $g_{k,q}$ with $q=\pi$
total transferred momentum will be considered.
Specifically we consider a toy model model consisting in a two-site lattice with
two electrons which can occupy a truncated Hilbert space made by the
following $S_z=0$ states:
$$
%\begin{equation}
|1\rangle=|c^{\dag \su }_{\pi}c^{\dag \gu}_{\pi} |0\rangle \ , \  
|2\rangle=|c^{\dag \su }_{0}c^{\dag \gu}_{0} |0\rangle  \ , \ 
|3\rangle=|a^{\dag}_{\pi}c^{\dag \su }_{0}c^{\dag \gu}_{\pi} |0\rangle  \ , \ 
|4\rangle=|a^{\dag}_{\pi}c^{\dag \gu }_{0}c^{\dag \su}_{\pi} |0\rangle  \ .
%\end{equation}
$$
The crucial simplification of this model is that the phonon Hilbert space is truncated
to the subspace with only zero or one phonon. Of course this is a severe limitation
if one aims to apply the model to the case of strong {\it e-ph}
couplings, but is acceptable, instead, for the general analysis of the method and for the
investigation of the weak {\it e-ph} coupling regime. 
We use the following single-electron dispersion
\begin{equation}
\epsilon_q=-2t \cos(q).
\end{equation}
In matricial representation (\ref{e_draft_0.5}) takes the form:
\begin{equation}
H(l)=
\left( 
\begin{array}{cccc}
\tilde{\epsilon}_{\pi}(l) & V_{0,0,\pi}(l)            & g_{\pi,\pi}(l)      & -g_{\pi,\pi}(l)\\
V_{0,0,\pi}(l)            & \tilde{\epsilon}_{0}(l)   & g_{0,\pi}(l)        & -g_{0,\pi}(l)  \\
g_{\pi,\pi}(l)            & g_{0,\pi}(l)              & \lambda(l)          & -V_{0,\pi,\pi}(l)  \\
-g_{\pi,\pi}(l)           & -g_{0,\pi}(l)             & -V_{0,\pi,\pi}(l)   & \lambda(l)     \\
\end{array}
\right);
\end{equation}
where
\begin{eqnarray}
\tilde{\epsilon}_{\pi}(l=0)&=2 \epsilon_{\pi}+U\\
\tilde{\epsilon}_{0}(l=0)&=2 \epsilon_{0}+U\\
\lambda(l=0)&=\epsilon_{\pi}+\epsilon_{0}+U+\omega_{0}.
\end{eqnarray}
Since we aim to eliminate the {\it e-e} interaction and see how the {\it e-ph} interaction changes,
we must implement a transformation to  eliminate the matrix elements (1,2), (2,1), (3,4) and (4,3) of $H$.
With the generator $\eta(l)$  given by Eq. \ref{e_draft_0.8} the flow equations read
\begin{eqnarray}
\frac{d\tilde{\epsilon}_{\pi}(l)}{dl}&=&
-4g_{0,\pi}(l)g_{\pi,\pi}(l)V_{0,0,\pi}(l)
+4g_{\pi,\pi}(l)^2 V_{0,\pi,\pi}(l) \nonumber \\
&+&2V_{0,0,\pi}(l)^2(\tilde{\epsilon}_{\pi}(l)-\tilde{\epsilon}_0(l))                  \\
\frac{d\tilde{\epsilon}_{0}(l)}{dl}&=&
-4g_{0,\pi}(l)g_{\pi,\pi}(l)V_{0,0,\pi}(l)
+4g_{0,\pi}(l)^2 V_{0,\pi,\pi}(l) \nonumber \\
&+&2V_{0,0,\pi}(l)^2(\tilde{\epsilon}_{0}(l)-\tilde{\epsilon}_{\pi}(l))                \\
\frac{d\lambda(l)}{dl}&=&
4g_{0,\pi}(l)g_{\pi,\pi}(l)V_{0,0,\pi}(l)
-2g_{0,\pi}(l)^2V_{0,\pi,\pi}(l)
-2V_{0,\pi,\pi}(l)g_{\pi,\pi}(l)^2                                                   \\
\frac{dV_{0,0,\pi}(l)}{dl} &=&
-(\tilde{\epsilon}_{0}(l)-\tilde{\epsilon}_{\pi}(l))^2V_{0,0,\pi}(l)
-2g_{0,\pi}(l)^2 V_{0,0,\pi}(l) \nonumber \\
&+&4g_{0,\pi}(l)g_{\pi,\pi}(l)V_{0,\pi,\pi}(l)
-2g_{\pi,\pi}(l)^2V_{0,0,\pi}(l)                                                     \\
\frac{dV_{0,\pi,\pi}(l)}{dl} &=& 
-2g_{0,\pi}(l)^2V_{0,\pi,\pi}(l)
+4g_{0,\pi}(l)g_{\pi,\pi}(l)V_{0,0,\pi}(l)
-2g_{\pi,\pi}(l)^2V_{0,\pi,\pi}(l)                                                   \\
\frac{dg_{\pi,\pi}(l)}{dl} &=&
g_{0,\pi}(l)V_{0,0,\pi}(l)(2\tilde{\epsilon}_{\pi}(l)-\tilde{\epsilon}_{0}(l)-\lambda(l))
+g_{\pi,\pi}(l)V_{0,\pi,\pi}(l)(-\tilde{\epsilon}_{\pi}(l)+\lambda(l)) \nonumber \\
&-& 2g_{0, \pi}(l)V_{0, 0, \pi}(l)V_{0,\pi,\pi}(l)
+g_{\pi,\pi}(l)V_{0,0,\pi}(l)^2 
+g_{\pi,\pi}(l)V_{0,\pi,\pi}(l)^2                                                    \\
\frac{dg_{0,\pi}(l)}{dl} &=&
+g_{0,\pi}(l)V_{0,\pi,\pi}(l)(-\tilde{\epsilon}_{0}(l)+\lambda(l))
+g_{\pi,\pi}(l)V_{0,0,\pi}(l)(-\tilde{\epsilon}_{\pi}(l)+2\tilde{\epsilon}_{0}(l)-\lambda(l)) 
\nonumber \\
&-&2g_{\pi, \pi}(l)V_{0, 0, \pi}(l)V_{0, \pi, \pi}(l)
+g_{0,\pi}(l)V_{0, 0,\pi}(l)^2
+g_{0,\pi}(l)V_{0,\pi,\pi}(l)^2                                                    
\end{eqnarray}
These equations can be solved numerically. As a consistency check we
 have compared the eigenvalues of both $H(l=0)$ and
$H(l=\infty)$ for different values of the starting couplings and
 obtained an exact match within 
our numerical precision.

In the many-site case we cannot deal with the full form of the flow
 equations. Therefore it is useful to test 
which approximation can be done without generating neither too big errors in the 
asymptotic values of the effective {\it e-ph} coupling nor a divergence in the 
case of degenerate levels. The simplest
choice is to make a weak-$U$ approximation stopping at first order in $U$.
Aiming to the effective {\it e-ph} interaction, we can neglect 
the flow equations for the energies because they would 
give higher order corrections when introduced in the flow-equations for the $g$'s:
\begin{eqnarray}
\frac{d\tilde{\epsilon}_{\pi}(l)}{dl}&=&
\frac{d\tilde{\epsilon}_{0}(l)}{dl}=\frac{d\lambda(l)}{dl}=0       \\
\frac{dV_{0,0,\pi}(l)}{dl} &=&
-(\tilde{\epsilon}_{0}(l)-\tilde{\epsilon}_{\pi}(l))^2V_{0,0,\pi}(l)
-2g_{0,\pi}(l)^2 V_{0,0,\pi}(l) \nonumber \\
&+&4g_{0,\pi}(l)g_{\pi,\pi}(l)V_{0,\pi,\pi}(l)
-2g_{\pi,\pi}(l)^2V_{0,0,\pi}(l)                                                     \\
\frac{dV_{0,\pi,\pi}(l)}{dl} &=&
-2g_{0,\pi}(l)^2V_{0,\pi,\pi}(l)
+4g_{0,\pi}(l)g_{\pi,\pi}(l)V_{0,0,\pi}(l) \nonumber \\
&-&2g_{\pi,\pi}(l)^2V_{0,\pi,\pi}(l)                                                   \\
\frac{dg_{\pi,\pi}(l)}{dl} &=&
g_{0,\pi}(l)V_{0,0,\pi}(l)(2\tilde{\epsilon}_{\pi}(l)-\tilde{\epsilon}_{0}(l)-\lambda(l))
\nonumber \\
&+&g_{\pi,\pi}(l)V_{0,\pi,\pi}(l)(-\tilde{\epsilon}_{\pi}(l)+\lambda(l))                 \\
\frac{dg_{0,\pi}(l)}{dl} &=&
g_{0,\pi}(l)V_{0,\pi,\pi}(l)(-\tilde{\epsilon}_{0}(l)+\lambda(l)) \nonumber \\
&+&g_{\pi,\pi}(l)V_{0,0,\pi}(l)(-\tilde{\epsilon}_{\pi}(l)+2\tilde{\epsilon}_{0}(l)-\lambda(l))  \\
\end{eqnarray}
We analyzed the behavior of the approximated effective {\it e-ph} couplings comparing them to the 
exact case for different values of $g_0$ and 
 $U\in [0,3], \ \ \omega_0 =0.02$. For the sake
of convenience, we have used energy units such that $t=1/2$.
As far as $\omega_0$ is concerned, we consider the most
frequent adiabatic regime with a typical value corresponding 
to a hundredth of the maximum electron bandwidth.

\begin{figure}[!ht]
\begin{center}
\includegraphics[angle=-90,scale=.50]{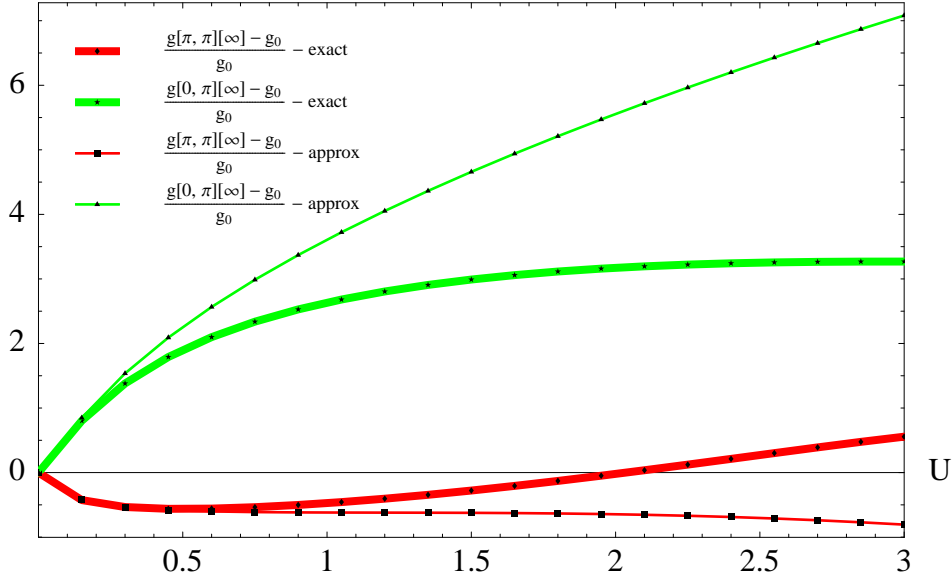}
\caption{Exact(thick) and approximate (thin) 
numerical results for $g_{0,\pi}$ (green) and $g_{\pi,\pi}$(red) vs U. 
Here we assume $g_{k,q}(l=0)=0.3$, $\omega_0=0.02$.}
\label{matricial_1}
\end{center}
\end{figure}

In figure (\ref{matricial_1}) we show a comparison between approximated and exact asymptotic values of
$g_{k,q}$ for various values of $U$. 
In this example strong renormalizations of $g_{k,q}$ at various $U$ are present, both in the
exact and in the approximated expressions. 
The approximated solutions are in good agreement with the exact
ones only for values of electronic repulsion U smaller than $t$.
Both these phenomena, the strong renormalizations of $g$ and the sizable differences
between the exact and the approximate results, are due to the big energy differences
between the states with and without phonons, basically controlled by 
the energy-level difference of order $t$ and to the
small value of $g_0$ we have been using. 
As far as the second point is concerned,
the approximated and the exact values are different because the vanishing of 
$V_{0,0,\pi}$ is controlled by $t$, whereas $g_0^2$ controls the vanishing
of $V_{0,\pi,\pi}$, which for small values of $g_0$ goes to zero slowly in the
flow equations. The presence of $g_0^2$ is necessary in the equation for $V_{0,\pi,\pi}$
to get an asymptotically vanishing solution. Otherwise, due to the degeneracy of
the levels $\lambda$, $V_{0,\pi,\pi}$ would not flow at all.
The quadratic terms in $U$, which are missing
in our approximated expressions, modify strongly in this discrete case the asymptotic
values of $g$ when $U$ is not very small. 

To further clarify the physical interpretation of the various terms appearing in the 
exact asymptotic form of the Hamiltonian, we rewrite 
in real space the {\it e-ph} interaction. By disregarding the zero-momentum phonons,
the {\it e-ph} coupling for a fixed value of the spin takes the form
\begin{equation}
\frac{(g_\delta A^{\dag} + \bar{g}_\delta A)\delta+
(\bar{g}_J A - g_J A^{\dag}  )i J}{2}
\end{equation}
where
$\delta= c^{\dag}_{1}c_{1}-c^{\dag}_{2}c_{2}, J= i(c^{\dag}_{1}c_{2}-c^{\dag}_{2}c_{1}), 
A= a_{1}-a_{2}
$
and
\begin{equation}
g_{\delta}=\frac{g_{0,\pi}+g_{\pi,\pi}}{2}, \,\,\,g_{J}=\frac{g_{0,\pi}-g_{\pi,\pi}}{2}.
\end{equation}
We explicitely checked that at very large $U$,
$g_{J}$ grows monotonically with $U$ while $g_{\delta}$ decreases
vanishing asymptotically in $U$.
This result naturally arises from the correlation-induced suppression of 
charge fluctuations. Intuitively, as $U$ increases the charge fluctuations
become stiffer and this reflects in a decrease of the {\it e-ph} coupling that
couples with the charge inbalance. Previous analyses already found that at  small
transferred momenta, the vertex corrections to $g$ arising from the {\it e-e} interactions
suppress the $e-ph$ coupling by a factor $(1+F_0^s)$, where $F_0^s$ is the standard
Landau parameter entering the compressibility and is an increasing function of $U$ 
\cite{grilli,zeyher}.
Indeed the out-of-phase phonon $A$ couples with the charge unbalance
between the two sites. Since charge-density fluctuations are hindered by the local repulsion
$U$, this is mirrored at large $U$ in the suppression of the related coupling $g_\delta$. 
On the other hand,
the flow equations implement a transformation, which obviously keeps the spectrum unchanged.
Since the energy scale $U$ is present in the starting Hamiltonian, the same scale must
appear in the transformed Hamiltonian as well. $g_\delta$ vanishes for large $U's$, and the
interacting system keeps the scale $U$ via
the dressed {\it e-ph} interaction $g_J$. In particular, in the large-$U$ limit (which can be
dealth with exactly in the $t=0$ limit) one can show
that the local repulsion $U$ is replaced by a nearest-neighbor attraction 
$V\sim g_J^2/\omega_0$.
This leads to the relation $g_J \sim \sqrt{\omega_0 U}$, which we numerically tested in the
large-U limit. 
The appearance of a non-local {\it e-ph} coupling is also obtained by a standard
Lang-Firsov trasformation
$$
\exp{S_{LF}}=\exp{-\frac{g}{\omega_0}\left[ - \sum_{i\sigma} n_{i\sigma} \left( a^\dagger_i-
 a_i\right)\right]}
$$
which is usually implemented 
to eliminate the linear local {\it e-ph} coupling in favor
of an effective local {\it e-e} attraction $-2g^2/\omega_0$
and a coupling between the phonons and the hopping term. 
By expanding to first order in the phonon fields this last term
one can easily check that a linear non-local {\it e-ph} coupling
$$g_J=- \frac{te^{-\alpha^2}\alpha}{2} \sum_{i,j,\sigma} \left( c_{i\sigma}^\dag c_{j\sigma} 
-c_{j\sigma}^\dag c_{i\sigma} \right)\left( a^\dag_i-a^\dag_j -(a_i-a_j) \right).
$$
is obtained. Then  the specific value $g=\sqrt{U\omega_0/2}$ also 
leads to the elimination of the {\it e-e} repulsion in the transformed Hamiltonian. Therefore
for this very specific choice, the effect of the Lang-Firsof transformation
is similar to the one generically obtained via our flow equations at large values of $U$, where
the local {\it e-ph} coupling $g_\delta$ vanishes together with $V$, while
$g_J(l=\infty)$ tends to $\sqrt{U\omega_0/2}$.
\footnote{We also find in the Lang-Firsov-generated {\it e-ph} coupling
a proportionality to the exponential factor $\exp[{-(g/\omega_0)^2}]$. This factor is
somewhat misleading since it should disappear in the effective four-fermion interaction
as a result of the summation over intermediate multi-phononic states (see, e.g., the 
effective superexchange coupling in the Hubbard-Holstein model of Ref. \cite{stephan}
or the effective Kondo coupling in the Anderson-Holstein model of Ref. \cite{cornaglia}).
Here the limitation of 0 or 1 phonon states  prevents this cancellation to occur.}

Before going on analyzing the 2D continuum case, we reassert
 the two relevant indications of this section: The first is that it is indeed possible
to eliminate the {\it e-e} interaction matrix elements obtaining a
convergent solution for the effective {\it e-ph} couplings. The second indication
is that the presence of an effective phonon-mediated interaction introduces
the quadratic terms in {\it e-ph} coupling in the flow equations of the {\it e-e}
coupling, allowing to get a convergent asymptotic solution
with a vanishing effective $V$ already at first order in $U$.

%%%%%%%%%%%%%%%%%%%%%%%%%%%%%%%%%%%%%%%%%%%%%%%%%%%%%%%%%%%%%%%%%%%%%%%%%%%%%%%%%%%%%%%%%%%%%%%

\subsection{Effective {\it e-ph} interaction}                             \label{ss_draft_4_2}
So far we discussed the generalities of the method and  showed the possibility 
to have a convergent
solution while eliminating some matrix elements in favor of others,
 using the generator (\ref{e_draft_0.8}) within our toy model. Here we proceed to the
calculation of the effective {\it e-ph} coupling in the more general case of a 
two dimensional continuum. In this work, for simplicity we
limit our analysis to the first order in the bare 
{\it e-e} coupling so that the flow equations are obtained from the following commutators   
\begin{eqnarray}
\frac{dg_{k,q}(l)}{dl}      &\to&   [[G,V],T] + [[T,V],G]  + o(U^2)   \nonumber\\
\frac{dV_{k,k',q}(l)}{dl}   &\to&   [[T,V],T] + [[G,V],G]  + o(U^2)    \nonumber\\
\frac{d\epsilon_{k}(l)}{dl} &\to&   [[G,V],G]              + o(U^2)   \nonumber\\
\frac{d\omega_{q}(l)}{dl}   &\to&   [[G,V],G]              + o(U^2).\nonumber
\end{eqnarray}
Since our aim is the determination of an effective
{\it e-ph} coupling while eliminating $U$ at first order, we do not consider
the flow equations for the energies because they would add higher order terms in $U$ 
when inserted in the equations for the $g$'s. Having neglected the $U^2$ terms, 
corresponds to the $t\gg U$ and $g^2/t \gg U$ conditions.
 Thus the flow equations for $g_{k,q}(l)$ and $V_{k,k',q}(l)$ read 
\begin{eqnarray}
\frac{dV_{k,k',q}(l)}{dl}&=&-\Delta\epsilon_{k,k',q}^2 \ V_{k,k',q}(l) + ``[[G,V],G]"    \label{e_draft_1}\\
\frac{dg_{k,q}(l)}{dl}&=&\frac{1}{N}\sum_p g_{p-q,q}(l) V_{k,p,q}(l)(n_{p-q}-n_{p})
[2\Delta\epsilon_{k,k',q}+\alpha_{k,q} ]            \label{e_draft_2}
\end{eqnarray}
where 
\begin{eqnarray}
\Delta\epsilon_{k,k',q} &\equiv &\epsilon_{k+q} + 
\epsilon_{k'-q} - \epsilon_{k} - \epsilon_{k}  \nonumber\\
\alpha_{k,q} &\equiv& \epsilon_{k+q} - \epsilon_{k} + \omega_0  \nonumber
\end{eqnarray}
and 
``$[[G,V],G]$'' stands for ``{\it part of flow equations coming from the operatorial term 
$[[G,V],G]$}'' and makes it very
difficult to get an analytical solution. Nevertheless, as in the toy model, this term
is essential for those values of $k=\bar{k}$,
$k'=\bar{k'}$ and $q=\bar{q}$ such that $\Delta\epsilon_{\bar{k},\bar{k'}\bar{q}}=0$:
For these momenta, without this
term, $V_{\bar{k},\bar{k'},\bar{q}}(l)$ would not flow to zero. 
and Eq.~\ref{e_draft_2} would be non convergent. 

The term $[[G,V],G]$ is cumbersome, but for our purposes it will be
enough to keep the simplest diagonal part and neglect additional non-diagonal ones
with one of the three momentum indeces of $V$ under summation
\begin{equation}   \label{e_draft_2.5}
``[[G,V],G]"\simeq -8\ g_0^2 \ V_{k,k',q}(l).
\end{equation}
Working at first order in $U$, we have also replaced $g(l)$ by its bare value $g_0$. 
The solution for $V_{k,k',q}(l)$ is then
\begin{equation}
V_{k,k',q}(l)= U \ e^{-\big[ \Delta\epsilon_{k,k',q}^2 +
8g_0^2 \big] l}. 
\end{equation}
Again we can substitute $g_{p-q,q}$ in the sum in Eq.
(\ref{e_draft_2}) with its initial value $g_0$ and solve it
\begin{equation}
g_{k,q}(l)=g_0\left(1-\frac{U }{N}\sum_p 
\frac{e^{-\big[ \Delta\epsilon_{k,p,q}^2 + 8g_0^2 \big] l}
-1}{\big[ \Delta\epsilon_{k,p,q}^2 + 8g_0^2 \big]}
(n_{p-q}-n_{p})[2\Delta\epsilon_{k,p,q}+\alpha_{p-q,q}]\right)
                    \nonumber       
\end{equation}
so that 
\begin{equation}g_{k,q}(\infty)= g_0
\left(1+U \tilde{\chi}(k,q,\omega_0)  \right).
\label{e_draft_7.5}
\end{equation}
where
\begin{equation}
\tilde{\chi}(k,q,\omega_0) \equiv
\frac{1}{N}\sum_p \frac{(n_{p-q}-n_{p})[2\Delta\epsilon_{k,p,q}+\alpha_{p-q,q}]}{\big[
\Delta\epsilon_{k,p,q}^2 + 8g_0^2 \big]}.      
\end{equation}
To start with we assume $\epsilon_{k+q} - \epsilon_{k} + \omega_0=0$ (elastic scattering)
so that the $k$ dependence in the asymptotic solution disappears
and $\tilde{\chi}(k,q,\omega_0)$ becomes $\hat{\chi}(q,\omega_0)$ 
\begin{equation}                    \label{4.10}
\hat{\chi}(q,\omega_0)\equiv \frac{1}{N}\sum_p \frac{\left( n_{p-q}-n_{p}\right)\left(
\epsilon_{p-q} -
\epsilon_{p}- \omega_0\right)}{\left(\epsilon_{p-q} -
\epsilon_{p}- \omega_0\right)^2+8g_0^2}.
\end{equation}
If one further neglects the $g_0^2$ in
the denominator  the asymptotic solution becomes
\begin{equation}                    \label{e_draft_7.75}
g_{q}(\infty)= g_{0}\left( 1 + U\chi(q,\omega_0) \right)=
g_{0}\left( 1 + \ \frac{U}{N} \ \sum_p \frac{n_{p-q}-n_{p}}{\epsilon_{p-q} -
\epsilon_{p}- \omega_0} \right).
\end{equation}
where $\chi(q,\omega_0)$ is the usual Lindhardt polarization bubble.
It is worth noting that in the limit of small $\omega_0$, since $\sum_p
\frac{n_{p-q}-n_{p}}{\epsilon_{p-q} - \epsilon_{p}}<0$, we have 
$g_{q}(\infty)<g_{0}$ for every value of the transferred momentum $q$. 
On the other hand, the general expression for $g$ allows also for positive
correstions to $g$ in the small-$q$ region. Moreover in the small-$\omega_0$ 
limit, since 
\begin{equation}
{\hat \chi}(0,0)=
\lim_{q \to 0}\lim_{\omega \to 0} \ \frac{1}{N}\sum_p 
\frac{n_{p-q}-n_{p}}{\epsilon_{p-q} -\epsilon_{p}-
\omega} = -\mathcal{N} 
\end{equation}
(here the density of states  $\mathcal{N}$ is the usual static small-$q$ limit  
of the Lindhardt function) one obtains $g=g_0(1-U{\mathcal N})$.
This result coincides with the static limit of perturbation theory at first order in $U$,
thereby supporting the reliability of our approximations.
This result also perturbatively agrees with the finding that $e-e$ interactions
suppress the $e-ph$ coupling by a factor $(1+F_0^s)$ \cite{grilli,zeyher}, 
where at this order
$F_0^s=\mathcal{N}U$ is the standard Landau parameter entering the compressibility.

\begin{figure}[!ht]
\begin{center}
\includegraphics[angle=-90,scale=.60]{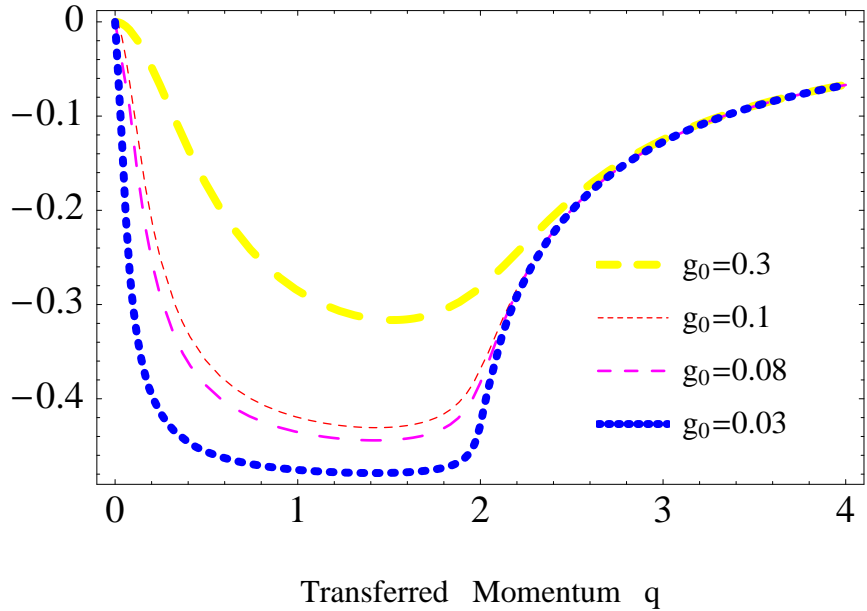}
\caption{The relative variation $\frac{g_{k,q}(\infty)-g_0}{g_0 \times U }$ of the {\it e-ph}
coupling. Here $E_{k,q}=\epsilon_{k+q}-\epsilon_{q}=0$, $\omega_0=0.02$. We recall that 
with our units, ${\mathcal{N}=1/2}$}.
\label{g_eff}
\end{center}
\end{figure}

Returning to the more general form of Eq. (\ref{e_draft_7.5}),
in figure \ref{g_eff} we show the behavior of the relative change of the effective 
coupling $g_{k,q}$. To simplify the numerical treatment we assume that electrons
have a free particle dispersion (we will also assume $\frac{\hbar^2}{2m}\equiv 1$): 
$
\epsilon_q=q^2 \ , 
$
and the direct self energy terms are included in the chemical potential
 $\mu$, which is taken to be unity. The constant density of states {\it per spin} 
is $\mathcal{N}=\frac{v}{4 \pi}$, where $v$ is the volume of the unit cell.
When one considers the case of one particle per site, $v=2\pi$.
Notice that, consistently with the assumption of a free-electron band dispersion,
no limit has been imposed to the bandwidth, thereby leading to a vanishing of the
``polarization-like'' functions ${\tilde \chi}$,  ${\hat \chi}$ and  ${\chi}$
in the limit of large transferred momenta $q \to \infty$.

The effective coupling is unchanged at $q=0$. This is due to the difference 
$(n_{p-q}-n_{p})$ which vanishes in the numerator 
and is not compensated by the denominator (as it would instead happen in the
Lindhardt function) because of the $g_0^2$ factor,
which makes the reduction of the {\it e-ph} coupling small at small $q$'s.
Conversely, upon increasing $q$ the reduction of $g$ becomes more pronounced.
This occurs up to $(v_Fq+ \omega_0)^2 \approx g_0^2$, when
the $\tilde{\chi}$ is well approximated by $\chi$. This latter in two dimensions, 
(if lattice effects are neglected), is constant up to $q=2k_F$ (and, with our units, 
$\chi =-{\mathcal{N}=-1/2}$) and then vanishes. As a result
$g$ reaches a plateau and finally recovers its bare value at large $q$'s.

The momentum dependence given in Fig. 3 shows the important feature
of the {\it e-ph} interaction in the presence of correlation of being peaked at
small momenta. The reduction of $g$ at large transferred momenta 
has been already found in previous treatments of the 
Hubbard-Holstein model with slave bosons \cite{kimtesanovic,grilli,pietronero}, Hubbard
operators \cite{zeyher} and quantum Monte Carlo \cite{scalapino}, where the momentum scale
at which $g$ decreses is set by $k_F$.  
Within a two-dimensional Monte Carlo approach \cite{scalapino}, it was found
that the decrease of $g$ is largest at intermediate $U$'s, and becomes smaller
at large $U$. Of course this non-monotonic result is not found 
with Eq. (\ref{e_draft_7.5}) here because of the weak coupling approximation.

\begin{figure}[!ht]
\begin{center}
%$\begin{array}{c@{\hspace{.5in}}c}
\includegraphics[scale=.6,angle=-90]{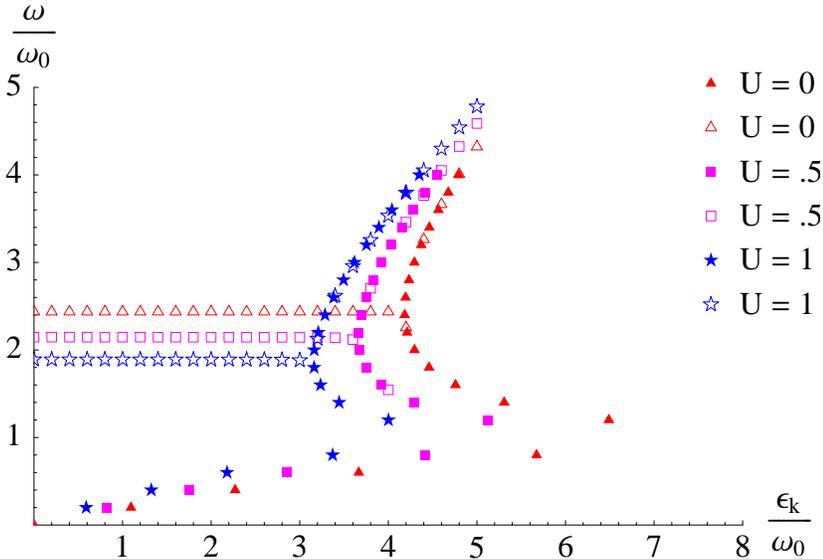} 
%\end{array}$
\caption{Dispersion of the peaks in the electronic spectral function
for different  values of U, and $g_0=0.3$, $\omega_0=0.02$.
We report $\hat{\omega}$ vs $\epsilon_k$ where ${\hat \omega}$ is the
solution of ${\hat \omega}-\epsilon_k=Re[\Sigma(k,{\hat \omega})]$ 
(filled symbols). We also report the position of the maximum of $A(k,\omega)$ for 
$\omega>\omega_0$ (unfilled symbols). }
\label{dispersion}
\end{center}
\end{figure}

Having obtained an effective {\it e-ph} interaction, which is 
reduced by the Hubbard repulsion $U$ and acquires a momentum dependence, it is of interest to
study its effects on the electronic dispersion. One should, however,
notice that the present first-order approximation  
limitates our possibility of obtaining substantial quantitative changes in the electronic
properties with respect to the standard Engelsberg-Schrieffer results in Ref. \cite{nglsbrg}.
We calculated the electron spectral density $A(k,\omega)=-(1/\pi) ImG(k,\omega)$,
where the electron Green's function contains the self-energy corrections at second order
in $g$. Fig. \ref{dispersion} displays the dispersions of the spectral-function peaks
for various values of $U$ at $\omega_0=0.02$ and $g_0=0.3$.
The overall behavior
is obviously similar to that found in \cite{nglsbrg}.
Specifically the {\it e-ph} coupling mixes the electron with the phonon and
bends the low-energy part of the dispersion on the phonon dispersion
at $E\sim \omega_0$. On the other hand at very large energy the electron
recovers its bare dispersion. 

It is clearly apparent that the {\it e-ph} interaction produces a break in the
electron dispersion, which is naturally more pronounced for larger {\it e-ph}
coupling (i.e. smaller $U$).
The presence of weak interaction $U$ introduces an
additional interesting feature:  Upon increasing $U$, 
the quantity $\lambda\propto g^2/(\omega_0t)$
is suppressed thereby reducing the electronic mass $m\approx m_0(1+\lambda)$ 
near the Fermi level. This is made visible in Fig.\ref{dispersion} by the increased velocity 
of the electrons at low energy. This decrease of the mass upon increasing the
interaction $U$ is a generic feature, which was also found within a DMFT treatment
of the Hubbard-Holstein model \cite{sangiovanni}.
We also notice that a difference in the low-energy and high-energy dispersions 
as well as the s-shaped dispersion of  Fig. 5,
(which displaces upon tuning the {\it e-e} interaction), are reminescent of the
kinks  \cite{KINKS,lnzr}  and of the breaks \cite{Sshape}
in the electronic dispersion of the cuprates experimentally observed along the $(1,\pm 1)$
directions and  around the $(\pm 1,0), (0,\pm 1)$  regions of the Brillouin zone 
respectively. These breaks are a rather generic feature of electrons coupled to
collective modes \cite{KINKS2,seibold,nagaosa,kulic}.

\section{Effective electron-electron interaction}     \label{s_draft_5}
An effective {\it e-e} coupling in the static limit can  be obtained from the
effective {\it e-ph} coupling of Eq. (\ref{e_draft_7.5}) considering the 
following two contributions diagrammatically depicted in Fig. 5
\begin{figure}[!ht]
\begin{center}
\includegraphics[angle=90,scale=.50]{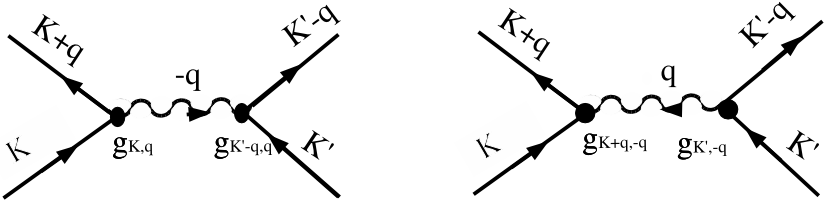}
\caption{e-e vertex from renormalized e-ph interaction.}
\label{U_eff}
\end{center}
\end{figure}

\begin{equation}
\label{ueff410}
V_{k,k',q}^{eff} = 
-\frac{g_{k,q}(l=\infty)g_{k'-q,q}(l=\infty)}{\omega_0}-
\frac{g_{k+q,-q}(l=\infty)g_{k',-q}(l=\infty)}{\omega_0}
\end{equation}
If one restricts the external momenta $k,k'$ 
to the domain $I$ such that the condition of energy conservation
$\epsilon_{k'} -\epsilon_{k'-q}=  \epsilon_{k+q} - \epsilon_{k}$
is satisfied, one obtains a simple expression for $V_{k,k',q}^{eff}$
\begin{eqnarray}
V_{k,k',q}^{eff} &=& 
-\frac{2g_0^2}{\omega_0} \nonumber \\
&-&\frac{4g_0^2U}{\omega_0N}\sum_p \frac{(n_{p-q}-n_{p})[2(\epsilon_{k+q} 
+ \epsilon_{p-q} - \epsilon_{p} - \epsilon_{k})+
(\epsilon_{p} - \epsilon_{p-q} )]}{ \big(\epsilon_{k+q} + \epsilon_{p-q} - \epsilon_{p}
- \epsilon_{k}\big)^2 + 8g_0^2} \\
\,\,\,\, k,k'  &\in &\,\,\, I \nonumber
\end{eqnarray}
In this way we obtain the first correction of order $U/t$ to the purely phononic
limit $U_{eff}\approx -2g_0^2/\omega_0$ \footnote{The use of the perturbative
diagrammatic approach in Fig. \ref{U_eff} implies the smallness of the {\it e-ph}
coupling $g_0\ll t$, since otherwise higher order phononic corrections (self-energy and
vertex) should be considered. It is worth noticing that this limitation is instead not
required in Section 4.2, where the elimination of the coupling $V$ with the neglect
of some commutators imposes conditions on the smallness of $V$ coupling only.}
We also notice that, imposing the elastic-scattering condition  
$\epsilon_{k'-q} - \epsilon_{k'} =  \epsilon_{k+q} - \epsilon_{k}=0$
one retrieves the perturbative result 
$
V_{q \ - \ elastic}^{eff} =-(2g_0^2/\omega_0)\left(
1+U{\hat \chi}(q,\omega_0)+U{\hat \chi}(q,-\omega_0)\right)=-(2g_0^2/\omega_0)\left(
1+2U{\hat \chi}(q,0)\right)
$
where ${\hat \chi}$ is the modified ``polarization'' bubble of Eq. (\ref{4.10}).

We have obtained a reduced attractive interaction modified according to 
the incipient effect of correlation
on the {\it e-ph} interaction. In order to get further insight on the 
interplay between electronic correlations and 
the {\it e-ph} interaction we turn to the study of 
the effective {\it e-e} coupling by reversing the previous procedure:
We use the flow equation technique to eliminate the {\it e-ph} 
interaction in favor of the effective $V(k,k',q)$. 
The generator of the unitary transformation is
\begin{equation}
\eta(l)= [:T:+:V:,:G:]
\end{equation}
leading to the flow equations
\begin{eqnarray}
\frac{dG}{dl} &\to&
[[:T:,:G:],:T:]+[[:T:,:G:],:V:] \nonumber \\
&+& [[:V:,:G:],:T:]+[[:V:,:G:],:V:] \\
\frac{dV}{dl} &\to&[[:T:,:G:],:G:]+ [[:V:,:G:],:G:] 
\end{eqnarray}
Here we do not consider the flow equations for the energies under the condition that
$(t,\omega_0)\gg g_0$. 
Moreover, when inserted in the equations for the
effective $V_{k,k',q}$, they would add higher-order corrections in $g_0$.
In the equation for $G$ we neglect terms of order $U^2g_0$ (the last commutator
in the r.h.s.). In the equation for $V$ we 
disregard terms of order $Ug_0^2$ (the last commutator in the r.h.s.). Both these
approximations are valid under the condition $t\gg U$. 

When rewritten in terms of couplings and with the above approximations, the 
flow equations read
\begin{eqnarray}                       \label{e_letter_4}
\frac{dg_{k,q}(l)}{dl}&=& -\alpha_{k,q}^{2} \ g_{k,q}(l) \nonumber \\
&-&\frac{1}{N}\sum_{p}g_{p-q,q}(l)V_{k,p,q}(l)(2\alpha_{p-q,q}+
\Delta\epsilon_{k,p,q})(n_{p-q}-n_{p}) \\
                                       \label{e_letter_5}
\frac{dV_{k,k',q}(l)}{dl}&=&-\left(\alpha_{k'-q,q}+\alpha_{k,q}\right) g_{k'-q,q}(l) \ g_{k,q}(l)
\nonumber \\
&-&\left(\alpha_{k+q,-q}+\alpha_{k',-q}\right) g_{k+q,-q}(l) g_{k',-q}(l)
\end{eqnarray}
where, again, $\alpha_{k,q}=\epsilon_{k+q}-\epsilon_{k}+\omega_0$, 
$\Delta\epsilon_{k,k',q}=\epsilon_{k+q}+\epsilon_{k'-q}-\epsilon_{k'}-\epsilon_{k}$,
and $\varepsilon_k\equiv -2t\left[\cos(k_x)+\cos(k_y)\right]$.
The $U=0$ case was previously considered in Ref. \cite{lnz}.
In this limit, Eq. (\ref{e_letter_4}) and
(\ref{e_letter_5}) can be solved without any further approximation yielding
\begin{equation}
g_{k,q}(l)= g_0 \ Exp\{-\alpha_{k,q}^{2} \ l \} 
\end{equation}
for the {\it e-ph} coupling and 
\begin{equation}                         \label{e_letter_6}
V_{k,k',q}(l=\infty)=
g_0^2\frac{-\alpha_{k'-q,q}-\alpha_{k,q}}{\alpha_{k'-q,q}^2+\alpha_{k,q}^2}
+g_0^2\frac{-\alpha_{k+q,-q}-\alpha_{k',-q}}{\alpha_{k+q,-q}^2+\alpha_{k',-q}^2}
\end{equation}
as the new effective {\it e-e} coupling. As in Ref. \cite{lnz}
Eq. (\ref{e_letter_6}) in the BCS channel($k'=-k$) gives
\begin{eqnarray}
V_{k,-k,q}(l=\infty) &=& -g_0^2\frac{2\omega_0}{2(\epsilon_{k}-\epsilon_{k+q})^2+2\omega_0^2}
-g_0^2\frac{2\omega_0}{2(\epsilon_{k}-\epsilon_{k+q})^2+2\omega_0^2} \nonumber \\
&=& -\frac{2g_0^2\omega_0}{(\epsilon_{k}-\epsilon_{k+q})^2+\omega_0^2}
\end{eqnarray}
while in the case of scattering with energy conservation
($\epsilon_{k+q}-\epsilon_{k}=-\epsilon_{k'-q}+\epsilon_{k'}$)  it gives
\begin{equation}
V_{elastic}(l=\infty)=-\frac{g_0^2}{\alpha_{k,q}}-\frac{g_0^2}{\alpha_{k+q,-q}}=
g_0^2\frac{2\omega_0}{(\epsilon_{k+q}-\epsilon_{k})^2-\omega_0^2}
\end{equation}
As discussed in Ref. \cite{lnz}, the effective interaction in the BCS channel is 
always attractive, with a noticeable difference with respect to the result of
Fr\"olich \cite{froelich}.

In the generic $U>0$ case the equations (\ref{e_letter_4}) and (\ref{e_letter_5})
are not analitically tractable and one should treat them numerically.
Here instead we aim to get an insight on the enegy scales involved in the
phonons dressing the effective {\it e-e} interaction and therefore we consider an
approximate treatment. We expect that the dependence of the
effective couplings on the momenta of the fermionic legs is less
important than the dependence on the transferred momentum. This is particularly
reasonable if one is mostly interested in the physically relevant region
of scattering processes around the Fermi surface.
According to this expectation, we assume here that $g_{k,q}$ mostly depends 
on the transferred momenta  $q$ and we neglect its $k$ dependence.  
Under this assumption the electronic parts of the $\alpha$'s in Eq.(\ref{e_letter_5})
cancel leaving $\omega_0$ as the only scale and the $k$ and $k'$ dependence of $V$ 
drops. With the further assumption that in Eq. (\ref{e_letter_4}) $V$ can be replaced by
its bare value $U$, Eqs. (\ref{e_letter_4}) and (\ref{e_letter_5}) become
\begin{eqnarray}
\frac{dg_{q}(l)}{dl} &=& -\langle \alpha_{k,q}^{2}\rangle g_{q}(l) - 
\frac{Ug_{q}(l)}{N}\sum_{p}(\epsilon_{p}-\epsilon_{p-q})(n_{p-q}-n_{p})
\label{averagedg}
\\
\frac{dV_{q}(l)}{dl}&=&-4\omega_0 g_q^2 (l) 
\label{averagedV} 
\end{eqnarray}
In deriving the equation for $g$ we exploited the obvious property 
$\sum_p(2\omega_0 + (\epsilon_{k}-\epsilon_{k-q})(n_{p-q}-n_{p})=0$.
Moreover to eliminate the $k$ dependence from the r.h.s., we consider the average
quantity  $\alpha_{k,q}^{2}  \approx \langle   \alpha_{k,q}^{2}\rangle $ over 
the momenta $k$ around the Fermi surface. The resulting expression for
$\langle \alpha^2 \rangle$ can be cast in the form
\begin{equation}
\langle\alpha_{k,q}^2 \rangle  = t^2 \beta_q+ \omega_0t\gamma_q + \omega_0^{2},
\end{equation}
where the functions $\beta_q$ and $\gamma_q$ depend on the specific average procedure
(e.g., from the width of the shell around the Fermi surface over which the average is
carried out). In particular, the average could be  taken strictly on the Fermi surface,
thereby getting explicit expressions. In  general one can notice that 
both $\beta_q$ and $\gamma_q$ vanish for $q\to 0$ and that $\langle\alpha_{k,q}^2 \rangle$
is obviously a non-negative quantity. 

We have not reported here the cumbersome
expressions of the $O(U^2g_0)$ and  $O(g_0^2U)$ terms, but we explicitely checked that
under the conditions $V_{k,p,q} \simeq U$, adopted in the
evolution for $g$ and $g_{k,q}(l) \sim g_{q}(l)$,
these additional terms vanish and no further corrections should be considered in the
flow equations. Therefore  within this approximation, the limitation of
small $U$ can be somewhat relaxed.

The solution of Eq. (\ref{averagedg}) is
\begin{equation}
g_{q}(l)= g_0 Exp\Big\{\Big[- (t^2\beta_q + t\omega_0 \gamma_{q} + \omega_0^{2}) -
 \frac{2tU}{N}\sum_{p}(\bar{\epsilon}_{p}-\bar{\epsilon}_{p-q})(n_{p-q}-n_{p}) 
\Big]l\Big\}
\end{equation} 
Where we use the notation ${\epsilon}_{p}=2t \bar{\epsilon}_{p}$.
Here we note that the exponential argument is manifestly negative and it implies 
the vanishing of $g_{q}(l)$ for $l\to\infty$. We can now calculate the effective 
electron-electron coupling
\begin{equation}                                           \label{veff-average-no-p-h}
V_{k,k',q}(l=\infty) = U - 
\frac{2g_0^2 (\omega_0)}
{t^2\beta_q+t\omega_0\gamma_q  + \omega_0^{2}+\frac{2tU}{N}\sum_{p}
(\bar{\epsilon}_{p}-\bar{\epsilon}_{p-q})(n_{p-q}-n_{p})}.
\end{equation}
The appearance of $\omega_0$ only in the numerator is a consequence of the above-mentioned
cancellation of the electronic scale in the $\alpha$'s strictly valid under the assumption
that $g_q$ does not depend on $k$.
The effective coupling has four non-negative contributions in the denominator, with
purely phononic, purely electronic and mixed character:
$\omega_0^2$, $t^2 \beta_q$, $t\omega_0\gamma_q$ and $2Ut\eta_q\equiv \frac{2Ut}{N}\sum_{p}
(\bar{\epsilon}_{p}-\bar{\epsilon}_{p-q})(n_{p-q}-n_{p})$ 
The last three expressions  vanish at $q=(0,0)$ and one finds in this limit 
$-2g_0^2/\omega_0$ as a correction to $U$. The presence of correlation 
generically reduces this attractive contribution to $V_{k,k',q}$ when $q$ is
finite. Depending on the relative importance of the various terms appearing in the
denominator of Eq. \ref{veff-average-no-p-h} as determined by the parameters $t$, $U$ 
and $\omega_0$, the effective interaction acquires different limiting forms. 
First of all we notice that the correct anti-adiabatic limit 
$(\omega_0 /t),(\omega_0/U) \to \infty$ is  recovered
\begin{equation} \label{antiad-lim}
V_{k,k',q}(l=\infty) =
U - \frac{2g_0^2}{\omega_0}.
\end{equation}
which was also obtained in the limit $q\to 0$.
When $\omega_0 \gg t$, a further momentum dependent correction $\gamma_q t/\omega_0$ is obtained.

In the opposite limit, ($\omega_0 \to 0$, but with $\lambda=2g_0^2/(\omega_0 \, t)$ finite)
we find that the whole contribution of the {\it e-ph}
attraction vanishes for $\omega_0\to 0$. This is reasonable, since an instantaneous
effective interaction cannot be affected by a static lattice deformation. 
We notice that in Ref. \cite{sangiovanni}, where the effective $U$ was also calculated
within the DMFT approach, a phonon correction linearly vanishing with $\omega_0$
was also found. Numerically the correction involved the scales $U\sim t $ and $\omega_0$.
An effective $U$ was also obtained by means of a slave-boson-Lang-Firsov 
treatment of the Hubbard-Holstein model \cite{barone}, where in the adiabatic limit
$t\gg \omega_0$, it was found $U_{eff}\approx U-g^2/4t$ for a square lattice in two dimensions.
In our case, in the same adiabatic limit, our Eq. (\ref{veff-average-no-p-h}) results in 
\begin{equation}
U_{eff}\approx U-\frac{2g^2}{t\beta_q+2U\eta_q}\frac{\omega_0}{t}.
\end{equation}
Our present crude approximate treatment therefore reproduces the results of the DMFT and of the
Lang-Firsov-Slave-Boson approaches with an additional $\omega_0/t$ factor. This factor
in this limit provides a substantial reduction of the phononic corrections at finite $q$'s,
while this reduction is less and less severe at small transferred momenta, owing to the
vanishing of $\beta_q$ and $\eta_q$ in this limit. On the other hand, going back to
Eq. (\ref{averagedV}), one should remember that relaxing the assumption of $k$ 
independence of $g$, an additional electronic scale coming from the $\alpha_{k,q}$ would appear
together with $\omega_0$. As long as $\omega_0$ is larger than $t$, these corrections are
unessential and the present approximation works well in the antiadiabatic limit. On the contrary,
in the adiabatic limit and finite $q$ these electronic corrections should appear in the numerator
of Eq. (\ref{veff-average-no-p-h}). As a consequence $U$ would be corrected by a term
of order $g^2/t$ without any additional $\omega_0/t$ factor.
This analysis therefore suggests  that the flow equations
 correctly capture the proper energy scales and  provide non-trivial momentum dependencies.

\section{Conclusions}                                 \label{s_draft_6}
This paper considered two main issues related to the interplay between {\it e-e}
correlations and {\it e-ph} coupling. We first investigated the effect of the {\it e-e}
Hubbard repulsion on the Holstein {\it e-ph} coupling, showing that this latter is generically
suppressed. This finding is consistent with previous results based on general
 Fermi-liquid arguments
\cite{grilli}, on large-N expansions based on the slave-boson 
\cite{grilli,kimtesanovic} or Hubbard-operator 
\cite{zeyher} techniques and on numerical quantum Monte Carlo approaches \cite{scalapino}
Remarkably, although we limited our discussion to the 
weak-correlation limit, we also find an agreement as far as the momentum dependencies are
concerned. Specifically we find that the Hubbard repulsion more severely suppresses the
{\it e-ph} coupling when large momenta are transferred.

 As already mentioned in the introduction this finding could explain the lack of
phonon features in transport properties (which mostly involve large transferred momenta) while
charge instabilities and $d$-wave superconductivity can arise from (mostly) forward scattering
and still be induced by a less suppressed {\it e-ph} coupling. 
The suppression of the {\it e-ph} coupling by {\it e-e} correlations is also
found within the Dynamical Mean-Field Theory
technique, which provides an exact treatment of the dynamics. However, although this technique
can be applied for any strength of the {\it e-e} and  {\it e-ph} couplings, it 
neglects the space correlations and provides poor informations on the momentum structure of the
renormalized couplings. In this sense our work based on the  flow-equation technique 
explores the correlation-induced renormalization of the {\it e-ph} coupling from a complementary 
point of view.

A second outcome of our research is provided by the elimination of the {\it e-ph} coupling
to obtain an effective {\it e-e} interaction. Remarkably our procedure, even though it has
been discussed within crude approximations, produces a reliable effective model for
correlated electron systems in the presence of the {\it e-ph} interaction. All the
energy scales, namely $\omega_0,g_0,t$ and $U$ are explicitely included and  
provide proper momentum dependence corrections in the different physical regimes
in agreement with  previous
analyses carried out with the DMFT and with the lang-Firsov-Slave-Boson approaches.

The phonons naturally introduce an effective
attractive interaction, which reduces the repulsive $U$. Again we find that the 
phonon-mediated interaction is more sizable at small transferred momenta. This allows to
infer some preliminary conclusions on the effects of phonons on the instabilities of the
Hubbard model. This issue had previously been addressed within the flow-equation technique
\cite{grt,hankevych}, where a phase diagram was proposed for the various instability lines
as a function of $U$ and doping $x$. Here we qualitatively remark that the phonons
induce an additional attraction, such that the effective repulsion $U(q)$ is smaller at 
smaller momenta. Therefore, the instabilities occurring at large momenta (like
antiferromagnetism or flux-phase) are only weakly affected by the phonon attractive part,
while the instabilities occurring at small momenta (phase separation, Pomeranchuk, superconductivity)
will occur at larger values of the bare repulsion parameter $U$.

{\it Acknowledgments} One of us (C.D.C.) gratefully acknowledges stimulating discussions with 
F. Wegner. We all acknowledge interesting discussions with M. Capone, C. Castellani, and
R. Raimondi. This work was financially supported by the Ministero Italiano dell'Universit\'a
e della Ricerca with the Projects COFIN 2003 (prot. $2003020230_006$ ) and COFIN 2005
(prot. $205022492_001$). 
One of us (C.D.C.) also thanks the von Humboldt Fundation.

% Bibliographic references with the natbib package:
% Parenthetical: \citep{Bai92} produces (Bailyn 1992).
% Textual: \citet{Bai95} produces Bailyn et al. (1995).
% An affix and part of a reference:
%   \citep[e.g.][Ch. 2]{Bar76}
%   produces (e.g. Barnes et al. 1976, Ch. 2).

\end{document}